\PassOptionsToPackage{hidelinks}{hyperref}

\documentclass[letterpaper,USenglish]{eptcs}

\usepackage[dvipsnames]{xcolor}

\usepackage{tikz}
  \usetikzlibrary{ 
  	cd,
  	math,
  	decorations.markings,
		decorations.pathreplacing,
  	positioning,
  	arrows.meta,
  	shapes,
		shadows,
		shadings,
  	calc,
  	fit,
  	quotes,
  	intersections,
    circuits,
    circuits.ee.IEC
  }
  
\usepackage{graphicx}
\graphicspath{{./figures/}}

\usepackage{amsmath}

\tikzset{
biml/.tip={Glyph[glyph math command=triangleleft, glyph length=.95ex]},
bimr/.tip={Glyph[glyph math command=triangleright, glyph length=.95ex]},
}
\tikzset{shorten <>/.style={shorten >=#1,shorten <=#1}}

\newcommand{\bimodfrom}[1][]{
	\begin{tikzcd}[ampersand replacement=\&, cramped]\ar[r, bimr-biml, "#1"]\&~\end{tikzcd}  
}

\title{Algebraic Property Graphs}
\author{Joshua Shinavier
\email{jshinavier@linkedin.com}
\institute{Uber Technologies, Inc.\footnote{For the toolkit, applications, and original formulation of APG. The author's current affiliation is LinkedIn Corporation.}}
\and
Ryan Wisnesky \qquad\qquad Joshua G. Meyers
\email{\quad ryan@conexus.com \quad\qquad jmeyers@conexus.com}
\institute{Conexus AI}
}
\def\titlerunning{Algebraic Property Graphs}
\def\authorrunning{J. Shinavier \& R. Wisnesky \& J. Meyers}

\usepackage[all,poly,matrix,arrow]{xy}

\usepackage{enumerate,makecell}

\usepackage{wrapfig}
\usepackage{mathtools}

\newtheorem{conjecture}{Conjecture}

\newcommand{\pgtype}{\textcolor{NavyBlue}}
\newcommand{\pgvalue}{\textcolor{RedViolet}}
\newcommand{\pgelement}{\textcolor{Bittersweet}}
\newcommand{\pglabel}{\textcolor{OliveGreen}}
\newcommand{\pglambda}{\pglabel{\boldsymbol{\lambda}}}
\newcommand{\pgsigma}{\pgtype{\boldsymbol{\sigma}}}
\newcommand{\pgsigmaprime}{\pgtype{\boldsymbol{\sigma'}}}
\newcommand{\pgtau}{\pgtype{\boldsymbol{\tau}}}
\newcommand{\pgupsilon}{\pgvalue{\boldsymbol{\upsilon}}}
\newcommand{\pgupsilonprime}{\pgvalue{\boldsymbol{\upsilon'}}}

\newcommand{\pgE}{\pgelement{\mathcal{E}}}
\newcommand{\pgV}{\pgvalue{\mathcal{V}}}
\newcommand{\pgT}{\pgtype{\mathcal{T}}}
\newcommand{\pgL}{\pglabel{\mathcal{L}}}
\newcommand{\pgLprime}{\pglabel{\mathcal{L'}}}

\newcommand{\pglabellit}[1]{\pglabel{\texttt{#1}}}
\newcommand{\pgtypelit}[1]{\pgtype{\texttt{#1}}}
\newcommand{\pgvaluelit}[1]{\pgvalue{\texttt{#1}}}
\newcommand{\pgzero}{\pgtype{0}}
\newcommand{\pgone}{\pgtype{1}}
\newcommand{\pgunit}{\pgvalue{()}}
\newcommand{\pgprimitives}{\pgtype{\sf P}}
\newcommand{\pgprimvals}{\pgvalue{\sf PV}}

\begin{document}

\maketitle


\begin{abstract}

We present a case study in applied category theory written from the point of view of an applied domain: the formalization of the widely-used \textit{property graphs} data model in an enterprise setting using elementary constructions from type theory and category theory, including limit and co-limit sketches. Observing that \textit{algebraic data types} are a common foundation of most of the enterprise schema languages we deal with in practice, for graph data or otherwise, we introduce a type theory for \textit{algebraic} property graphs wherein the types denote both algebraic data types in the sense of functional programming and join-union E/R diagrams in the sense of database theory. 
We also provide theoretical foundations for graph transformation along schema mappings with by-construction guarantees of semantic consistency. 
Our data model originated as a formalization of a data integration toolkit developed at Uber which carries data and schemas along composable mappings between data interchange languages such as Apache Avro, Apache Thrift, and Protocol Buffers, and graph languages including RDF with OWL or SHACL-based schemas.

\end{abstract}

\section{Introduction}

The notion of a \textit{property graph} originated in the early 2000s with the Neo4j\footnote{\url{https://neo4j.com}} graph database, and was popularized by what is now Apache TinkerPop,\footnote{\url{http://tinkerpop.apache.org}} a suite of vendor-agnostic graph database tools including the Gremlin graph programming language.
For most of their history, property graphs have been the stock-in-trade of software developers creating applications loosely based on variously mathematical notions of {\it labeled graph}, but with little formal semantics or type checking associated with the labels.
In that respect, property graphs differ from more heavyweight standards designed for knowledge representation, including the Resource Description Framework (RDF) and the Web Ontology Language (OWL).

Over the past decade, the developer community has increasingly turned to property graphs for large-scale data integration efforts including \textit{enterprise knowledge graphs}, i.e. graph abstractions that integrate a broad swath of a company's data, often drawn from a variety of internal data sources and formats.
These abstractions have expanded the de-facto meaning of graphs and have stretched the simple, intuitive property graph concept to its limits, leading to a shared sense of urgency around unifying abstractions for graphs~\cite{sakr2021future}, as well as recent community efforts around standardization, including GQL (Graph Query Language).\footnote{\url{https://www.gqlstandards.org}}
The authors of this paper are also involved in GQL, and our formalism was designed with an eye toward informing the emerging standard with respect to type systems for property graphs.
At the same time, the formalism described here was driven by enterprise use cases at Uber which demanded a particularly strict and precise notion of \textit{schema}.

We have chosen to define the semantics of our algebraic property graph formalism using \textit{category theory}, which emphasizes compositionality and abstract structure, and also comes equipped with a rich body of results about algebraic data types~\cite{mitchell}.  
Our use of category theory has also allowed us to implement this entire paper as a built-in example program in the open-source categorical query language CQL,\footnote{\url{http://categoricaldata.net}} which has close connections to algebraic databases.




This paper is organized as follows.
In \autoref{section.related}, we describe property graphs and other graph and non-graph data models which have been relevant to this work.
In \autoref{section.data-model}, we define algebraic property graphs as a data model and a formal logic, followed by examples in \autoref{section.examples}, connections to additional formalisms in \autoref{section.connections} and possible future extensions in \autoref{section.extensions}.
We briefly study transformations of algebraic property graphs along schema mappings in \autoref{sec:ops}, as well as an exploratory connection to Poly in \autoref{section.poly}.
Finally, in \autoref{section.implementations} we describe an industrial application of APG before we conclude in \autoref{section.summary}.


\section{Graph and non-graph data models}
\label{section.related}

For the sake of conciseness, we will focus only on property graphs in this paper, leaving aside the details of other data models we have used in practice, and mappings between those models and property graphs.
However, we will provide a high-level summary in the following.

\subsection{Property graphs}

\textit{Property graphs} are a family of graph data models which are typically concerned only with graph \textit{structure}; the \textit{semantics} are left to the application.
Every \textit{graph} in these data models is made up of a set of \textit{vertices} (or ``nodes'') connected by a set of directed, labeled \textit{edges} (or ``relationships'').
Vertices and edges are collectively known as \textit{elements}.
Every element has a unique identifier, and may be annotated with any number of key-value pairs called \textit{properties}.\footnote{In some cases, properties are also considered to be elements, as they are in APG.}

Beyond these basic commonalities, property graph data models start to differ.
Apache TinkerPop, which is the most widely used vendor-neutral property graph framework, allows graph data models to vary according to a number of ``features'' such as:
\begin{itemize}
\item Which primitive types are supported in the graph, and whether complex types such as lists, maps, and sets are supported
\item Which types, such as strings, integers, or UUIDs, may be used as unique identifiers
\item Whether vertex properties, edge properties, and/or meta-properties (all discussed below in \autoref{section.taxonomy}) are supported
\end{itemize}
Additional, vendor-specific schema frameworks provide further degrees of freedom that deal with such concepts as unlabeled, singly-labeled, and multiply-labeled vertices, inheritance relationships among labels, type constraints and cardinality constraints on properties and edges, higher-order edges, and so on.

\subsection{Resource Description Framework}

RDF~\cite{world2014rdf} is the primary W3C recommendation for knowledge representation on the Web.
RDF \textit{statements} are subject-predicate-object triples, any set of which forms an \textit{RDF graph}.
These graphs can be serialized in many formats, from XML-based formats to JSON-based and idiomatic text-based ones.
A basic schema language, RDF Schema, and a pattern-matching query language, SPARQL, are provided with RDF, both with a formal set-theoretic semantics.
These formal semantics represent a major advantage over existing property graph frameworks with respect to building enterprise knowledge graphs.  However, the complexity of the standards, together with structural limitations such as the lack of an agreed-upon way to express metadata about statements, are often cited as stumbling blocks, and have led to efforts to simplify RDF, enhance it, and/or reconcile it with the property graph data model.

The Dragon toolkit described in \autoref{section.implementations} supports two alternative schema languages: the Web Ontology Language (OWL) \cite{owl2012owl} and the Shapes Constraint Language (SHACL) \cite{knublauch2017shapes}.
The former is designed for knowledge representation, while the latter focuses on graph structure as opposed to semantics, and has been a particularly good fit for APG.

\subsection{Hypergraphs}
\label{section.hypergraph-data-models}

Although there are many notions of {\it hypergraph} in the computer science literature, the term usually refers either to a data structure which embodies the usual mathematical notion of a hypergraph, i.e. a graph in which a given edge may join any number of vertices, or a data structure in which edges are also vertices, and may be connected by further edges.
Hypergraph databases commonly combine these two features along with a notion of edge and/or vertex label, as well as labels for fields or roles, i.e. the named components of a hyperedge.
Some examples of hypergraph data models with concrete database implementations are Hypernode \cite{levene1990hypernode}, Groovy \cite{levene1991object}, HypergraphDB \cite{iordanov2010hypergraphdb}, and Grakn \cite{messina2017biograkn}.
None of these data models have been directly integrated with APG, although they provide conceptual reference points.



\subsection{Data exchange languages}
\label{section.data-serialization-languages}

At Uber, there are relatively few datasets which are explicitly ``graphs'', yet much of the company's data has the characteristics of a graph: labeled entities, with unique identifiers, connected by labeled relationships.
Schemas are expressed in a variety of languages, a few of which are illustrated in \autoref{table.rpc-comparison}.
\begin{figure}
\begin{center}
\footnotesize
\begin{tabular}{llllll}
& \textbf{\begin{tabular}[c]{@{}l@{}}Apache\\ Thrift\end{tabular}}
& \textbf{\begin{tabular}[c]{@{}l@{}}Apache\\ Avro\end{tabular}}
& \textbf{\begin{tabular}[c]{@{}l@{}}Protocol\\ Buffers v3\end{tabular}}
& \textbf{\begin{tabular}[c]{@{}l@{}}GraphQL\\ SDL\end{tabular}} \\
\hline
\textbf{product types} & yes & yes & yes & yes \\
\textbf{sum types}     & yes & yes & yes & yes \\
\textbf{interfaces}    &     &     &     & yes \\
\textbf{enumerations}  & yes & yes & yes & yes \\
\textbf{optionals}     & yes & yes &     & yes \\
\textbf{typedefs}      & yes &     &     &     \\
\textbf{defaults}      & yes & yes & yes & yes \\
\textbf{constants}     & yes &     &     &     \\
\textbf{lists/arrays}  & yes & yes & yes & yes \\
\textbf{maps}          & yes & yes & yes &     \\
\textbf{sets}          & yes &     &     &     \\
\end{tabular}
\end{center}
\caption{Comparison of selected data exchange languages}
\label{table.rpc-comparison}
\end{figure}
What most of these languages have in common is a system, usually unstated and informal, of algebraic data types.
This commonality made it possible to define APG as a shared logical data model, or what has been called a \textit{universal meta-model}~\cite{melnik}, and to translate data and schemas from one satellite model to another.

Of course, there are also numerous incompatibilities, and much of the work of defining information-preserving mappings between these languages, in practice, has to do with reconciling feature mismatches in a general-purpose way.
For example, all of the languages shown in \autoref{table.rpc-comparison} have an equivalent notion of \textit{product types}, instances of which are variously called ``records'', ``structs'', ``messages'', and so on.
Similarly, they all have a notion of \textit{sum types}, though in certain cases these have the semantics of a discriminated union (a \textit{coproduct}~\cite{awodey2010ct}), and in other cases, that of a set union.


\subsection{Relational databases}
\label{sec:relational}
There is a great deal of interplay between graph processing and relational database theory, and a correspondingly large amount of past research.
Here, we will only make some basic observations.
For example, a graph with directed edges and at most one edge between any given pair of vertices is equivalent to a binary relation: the edge relation of the graph.
Hence, we can encode such graphs and operations on them in SQL; generalizations of this encoding appear in many software systems.
Such encodings can also be used to prove inexpressivity results, including the result that no relational algebra query can compute the transitive closure of a graph's edge relation~\cite{Doan:2012:PDI:2401764}.
In practice, and despite these inexpressivity results, much graph processing is done on relational systems, and vice versa.

\section{Algebraic Property Graphs}
\label{section.data-model}


In this section, we formally define algebraic property graphs (APGs).
From this point on, a familiarity with category theory is required, and familiarity with database theory~\cite{Doan:2012:PDI:2401764} is helpful.
If this paper is rendered in color, the reader will see distinct colors for the \pglabel{labels}, \pgtype{types}, \pgelement{elements}, and \pgvalue{values} of graphs, concepts which will be described below.
The colors are intended to enhance readability, but are not essential to understanding the text.

The definition of an APG is parameterized by a set ${\pgprimitives}$, the members of which we call {\it primitive types},\footnote{We make no assumptions about the intended semantics of primitive types, which will typically encompass atomic objects such as booleans, character strings, integers, floating-point numbers, etc. However, the formalism can be extended to include complex objects, such as lists, trees, function abstractions in the sense of $\lambda$-calculus, built-in functions, and even objects in object-oriented programming notation, as described in \autoref{section.moretypes}.} and for each primitive type $\pgtype{p}$ in ${\pgprimitives}$, a set ${\pgprimvals}(\pgtype{p})$, the members of which we call the {\it primitive values} of type  $\pgtype{p}$.
First we define an {\it APG schema}. This consists of
\begin{itemize}
    \item a set $\pgL$, the members of which we call the \pglabel{{\it labels}}\footnote{e.g. $\pglabellit{User}$ or $\pglabellit{name}$}; and,
    \item a function $\pgsigma:\pgL\to \pgtype{\textrm{Ty}}(\pgL)$, providing the \pgtype{{\it schema}}\footnote{e.g. $\pgsigma(\pglabellit{name})$ $:= \pglabellit{User} \times \pgtypelit{String}$} of each label, where we define types $\pgtype{t}\in\pgtype{\textrm{Ty}}(\pgL)$ as terms in the following grammar:
\begin{align*}
\pgtype{t} \ ::= \ \pgzero \ & | \ \pgone \ | \ \pgtype{t_1} + \pgtype{t_2} \ | \ \pgtype{t_1} \times \pgtype{t_2} \
 | \ {\sf Prim} \ \pgtype{p} \ (\pgtype{p} \in {\pgprimitives}) \ | \ {\sf Lbl} \ \pglabel{l} \ (\pglabel{l} \in \pgL)
\end{align*}
\end{itemize}

We may omit writing {\sf Prim} and {\sf Lbl} when they are clear from the context.
Given an APG schema $S=(\pgL,\pgsigma)$, we then define an {\it APG on} $S$ as
\begin{itemize}
    \item a function $\pgelement{E}:\pgL\to\mathsf{Set}$ sending each label to its set of \pgelement{\it elements};\footnote{which we will indicate with subscripted symbols like $\pgelement{u_1}\in \pgelement{E}(\pglabellit{User})$ or $\pgelement{p_1}\in \pgelement{E}(\pglabellit{Place})$} and,
    
    \item for each $\pglabel{l}\in\pgL$, a function $\pgupsilon_{\pglabel{l}}:\pgelement{E}(\pglabel{l})\to \pgvalue{V}(\pgsigma(\pglabel{l}))$ sending each element to its \pgvalue{\it value}\footnote{e.g. $\pgupsilon_{\pglabellit{Name}}(\pgelement{n_1}) := (\pgelement{u_1}, \pgvaluelit{"Arthur"})$}, where the function $\pgvalue{V}:\pgtype{\textrm{Ty}}(\pgL)\to\mathsf{Set}$, which sends each type to its set of values,\footnote{e.g. $\pgvaluelit{42} \in \pgvalue{V}(\pgtypelit{Integer})$ or $(\pgelement{u_1},
    \pgvaluelit{"Arthur"}) \in \pgvalue{V}(\pglabellit{User} \times  \pgtypelit{String}$)} is defined recursively as follows:
    \begin{equation}\label{VintermsofE}
    \begin{aligned}
&\pgvalue{V}(\pgzero) := \pgzero \ \ \
\pgvalue{V}(\pgone) := \pgone \\
& \pgvalue{V}({\sf Prim} \ \pgtype{p}) := {\pgprimvals}(\pgtype{p}) \ \ \ 
\pgvalue{V}({\sf Lbl} \ \pglabel{l}) := \pgelement{E}(\pglabel{l}) \\ 
& \pgvalue{V}(\pgtype{t_1} + \pgtype{t_2}) := \pgvalue{V}(\pgtype{t_1}) + \pgvalue{V}(\pgtype{t_2})   \ \ \
\pgvalue{V}(\pgtype{t_1} \times \pgtype{t_2}) := \pgvalue{V}(\pgtype{t_1}) \times \pgvalue{V}(\pgtype{t_2})
\end{aligned}
\end{equation}
\end{itemize}

\subsection{Alternative formulation}

We can also think of APGs as type theories.
The following formulation makes the relationships between elements, labels, values, and types particularly easy to appreciate visually.
Given the APG schema $S=(\pgL,\pgsigma)$, an APG consists of:
\begin{itemize}
    \item A set $\pgE$ of elements.
    \item A labelling function $\pglambda: \pgE\to\pgL$ sending each element to its unique label.
    \item The set $\pgV$ of typed values, which are terms in the following grammar:
    \begin{align*}
\pgvalue{v} : \pgtype{t} \ ::= \ \pgunit : \pgone \ & | \ {\sf inl}_{\pgtype{t_2}}(\pgvalue{v} {:} \pgtype{t_1}) : \pgtype{t_1} + \pgtype{t_2} \ | \ {\sf inr}_{\pgtype{t_1}}(\pgvalue{v} {:} \pgtype{t_2}) : \pgtype{t_1} + \pgtype{t_2} \ \ | \ \ (\pgvalue{v_1} {:} \pgtype{t_1}, \pgvalue{v_2} {:} \pgtype{t_2}) : \pgtype{t_1} \times \pgtype{t_2} \\
& | \ {\sf Prim} \ {\pgvalue{v}}_{\pgtype{t}} : \pgtype{t} \ \ (\pgtype{t} \in {\pgprimitives}, \pgvalue{v} \in {\pgprimvals}(\pgtype{p})) \ \ \ 
 | \ \ \ {\sf Elmt} \ \pgelement{e} : \pglambda(\pgelement{e}) \ \ (\pgelement{e} \in \pgE) 
\end{align*}
We define the typing function $\pgtau:\pgV\to\pgT$ (where $\pgT\coloneqq \pgtype{\textrm{Ty}}(\pgL)$) by $\pgtau(\pgvalue{v} : \pgtype{t}):=\pgtype{t}$.
\item A function $\pgupsilon:\pgE\to\pgV$ sending elements to their values.
\end{itemize}

\noindent
We then impose the equation
$
\pgtau \circ \pgupsilon = \pgsigma \circ \pglambda
$
which states that the type of the value of each element is the same as the schema of the label of the element, ensuring that the structure of a graph always matches its schema.
This equation can be visualized as a commutative square:
$$
\xymatrix{
 & \pgE \ar[d]_{\pgupsilon} \ar[r]^{\pglambda} & \pgL \ar[d]^{\pgsigma} \\
 & \pgV \ar[r]^{\pgtau} & \pgT.
}
$$
This data determines an APG on $S$ by setting $\pgelement{E}(\pglabel{l}):=\pglambda^{-1}(\pglabel{l})$ and $\pgupsilon_{\pglabel{l}}:=\pgupsilon|_{\pgelement{E}(\pglabel{l})}$.

Conversely, given an APG on $S$ we can obtain a type theory of this kind by setting 
\begin{align*}
\pgE &:=\bigsqcup_{\pglabel{l}\in\pgL}\pgelement{E}(\pglabel{l})
\ \ \ \
\pglambda(\pgelement{e}\in \pgelement{E}(\pglabel{l})) := \pglabel{l}
\ \ \ \
\pgupsilon(\pgelement{e}\in \pgelement{E}(\pglabel{l})) := \pgupsilon_l(\pgelement{e})
\end{align*}

\section{Examples}
\label{section.examples}

\subsection{A generic graph}
\label{section.genericgraph}

\newcommand{\rdfpred}[1]{\texttt{#1}}

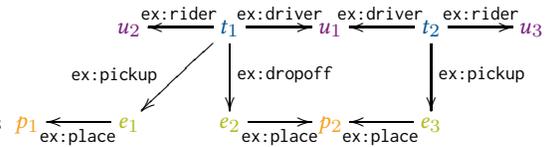
\begin{figure}
\begin{align*}
\xymatrixcolsep{2pc}\xymatrix{
&\pgelement{u_2}&\pgelement{t_1}\ar[l]_{\rdfpred{rider}}\ar[r]^{\rdfpred{driver}}\ar[dl]_>>>>>{\rdfpred{pickup}}\ar[d]^{\rdfpred{dropoff}}&\pgelement{u_1}&\pgelement{t_2}\ar[r]^{\rdfpred{rider}}\ar[l]_{\rdfpred{driver}}\ar[d]^{\rdfpred{pickup}}&\pgelement{u_3}& \\
\pgelement{p_1}&\pgelement{e_1}\ar[l]^{\rdfpred{place}}&\pgelement{e_2}\ar[r]_{\rdfpred{place}}&\pgelement{p_2}&\pgelement{e_3}\ar[l]^{\rdfpred{place}}& 
}
\end{align*}
\caption{An example graph}
\label{figure.rdf-graph}
\end{figure}

We will start with an example based on a simplified ride-sharing schema from Uber.
See \autoref{figure.rdf-graph} for a partial view of the example graph.
Let the schema of the APG be defined as follows:
\begin{footnotesize}
\begin{align*}
& \pgL := \{ \pglabellit{User}, \pglabellit{Trip}, \pglabellit{PlaceEvent}, \pglabellit{Place} \} \\
& \pgsigma(\pglabellit{User}) := \pgsigma(\pglabellit{Place}) := \pgone \\
& \pgsigma(\pglabellit{PlaceEvent}) := \pglabellit{Place} \times \pgtypelit{Integer} \\
& \pgsigma(\pglabellit{Trip}) := \pglabellit{User} \times \pglabellit{User} \times (\pgone + \pglabellit{PlaceEvent}) \times (\pgone + \pglabellit{PlaceEvent})
\end{align*}
\end{footnotesize}
The four labels of this particular APG are enumerated, and their schemas are given.
Note that the expression $\pgtype{1}+\pgtype{t}$ can be interpreted as optionality; some \normalfont{\pglabellit{Trip}}s have associated \normalfont{\pglabellit{PlaceEvent}}s, while others may not.
This is an example of a hypergraph rather than a typical property graph, as the schema of the \normalfont{\pglabellit{Trip}} label is more than binary.
Now an APG on this schema:
\begin{footnotesize}
\begin{align*}
& \pgE := \{ \pgelement{u_1}, \pgelement{u_2}, \pgelement{u_3}, \pgelement{t_1}, \pgelement{t_2},   \pgelement{e_1}, \pgelement{e_2}, \pgelement{e_3}, \pgelement{p_1}, \pgelement{p_2}, \pgelement{p_3} \} \\
& \pglambda(\pgelement{u_1}) := \pglambda(\pgelement{u_2}) := \pglambda(\pgelement{u_3}) := \pglabellit{User}
\ \ \ \
\pglambda(\pgelement{t_1}) := \pglambda(\pgelement{t_2}) := \pglabellit{Trip} \\
& \pglambda(\pgelement{p_1}) := \pglambda(\pgelement{p_2}) := \pglambda(\pgelement{p_3}) := \pglabellit{Place} \\
& \pglambda(\pgelement{e_1}) := \pglambda(\pgelement{e_2}) := \pglambda(\pgelement{e_3}) := \pglabellit{PlaceEvent} \\
& \pgupsilon(\pgelement{u_1}) := \pgupsilon(\pgelement{u_2}) := \pgupsilon(\pgelement{u_3}) := \pgupsilon(\pgelement{p_1}) := \pgupsilon(\pgelement{p_2}) := \pgupsilon(\pgelement{p_3}) := \ \pgunit \\
& \pgupsilon(\pgelement{t_1}) := ({\pgelement{u_1}}, {\pgelement{u_2}}, {\sf inr}({\pgelement{e_1}}), {\sf inr}({\pgelement{e_2}})) 
\ \ \ \
\pgupsilon(\pgelement{t_2}) := ({\pgelement{u_1}}, {\pgelement{u_3}}, {\sf inr}({\pgelement{e_3}}), {\sf inl}(\pgunit)) \\
& \pgupsilon(\pgelement{e_1}) := ({\pgelement{p_1}}, \pgvaluelit{1602203601})
\ \ \ \
\pgupsilon(\pgelement{e_2}) := ({\pgelement{p_2}}, {\pgvaluelit{1602203948}}) \\
& \pgupsilon(\pgelement{e_3}) := ({\pgelement{p_2}}, {\pgvaluelit{1602204122}})
\end{align*}
\end{footnotesize}
These elements and values capture the structure of the graph illustrated in \autoref{figure.rdf-graph}, as well as simple data such as timestamp values which are not depicted in the figure.
An interpretation of this APG as a typical property graph is that the elements $\{\pgelement{u_1}, \pgelement{e_2}, ...\}$ are not hyperelements, but rather vertices, with the projections between vertices as edges and the projections to data values as properties.
The following section will provide an example in which both vertices and edges are elements in the sense of APG.

\subsection{Taxonomy of property graph elements}
\label{section.taxonomy}

Typical property graphs -- i.e. those property graphs found in mainstream applications such as Apache TinkerPop -- are characterized by elements which are binary or less, and which limit the depth to which element references may be nested; they are a slight refinement of the classic graph data structure in which vertices are simple elements, and edges are pairs of vertices.

Let us revisit the ride-sharing example from \autoref{section.genericgraph}, but refactor the schema so that it becomes a typical property graph.
The schema will be as follows:
\begin{footnotesize}
\begin{align*}
& \pgL := \{ \pglabellit{User}, \pglabellit{Trip}, \pglabellit{Place}, \pglabellit{rider}, \pglabellit{driver}, \pglabellit{pickup}, \pglabellit{dropoff}, \pglabellit{pickupTime}, \pglabellit{dropoffTime} \} \\
& \pgsigma(\pglabellit{User}) := \pgsigma(\pglabellit{Trip}) := \pgsigma(\pglabellit{Place}) := \pgone \\
& \pgsigma(\pglabellit{rider}) := \pglabellit{Trip} \times \pglabellit{User} \\
& \pgsigma(\pglabellit{driver}) := \pglabellit{Trip} \times \pglabellit{User} \\
& \pgsigma(\pglabellit{pickup}) := \pglabellit{Trip} \times \pglabellit{Place} \\
& \pgsigma(\pglabellit{dropoff}) := \pglabellit{Trip} \times \pglabellit{Place} \\
& \pgsigma(\pglabellit{pickupTime}) := \pglabellit{pickup} \times \pgtypelit{Integer} \\
& \pgsigma(\pglabellit{dropoffTime}) := \pglabellit{dropoff} \times \pgtypelit{Integer}
\end{align*}
\end{footnotesize}
It is left as an exercise for the reader to compare this schema with that of \autoref{section.genericgraph} and observe that the generic graph is extremely similar, but more restricted (in useful ways); the second example has additional degrees of freedom which would require additional constraints, outside of the APG schema itself, to take away.
The following APG is analogous to the previous example:
\begin{footnotesize}
\begin{align*}
& \pgE := \{ \pgelement{u_1}, \pgelement{u_2}, \pgelement{u_3}, \pgelement{t_1}, \pgelement{t_2},   \pgelement{p_1}, \pgelement{p_2}, \pgelement{p_3}, \pgelement{r_1}, \pgelement{r_2}, \pgelement{d_1}, \pgelement{d_2}, \pgelement{i_1}, \pgelement{i_2}, \pgelement{o_1}, \pgelement{s_1}, \pgelement{s_2}, \pgelement{s_3} \} \\
& \pglambda(\pgelement{u_1}) := \pglambda(\pgelement{u_2}) := \pglambda(\pgelement{u_3}) := \pglabellit{User}
\ \ \ \
\pglambda(\pgelement{t_1}) := \pglambda(\pgelement{t_2}) := \pglabellit{Trip} \\
& \pglambda(\pgelement{p_1}) := \pglambda(\pgelement{p_2}) := \pglambda(\pgelement{p_3}) := \pglabellit{Place} \\
& \pglambda({\pgelement{r_1}}) := \pglambda({\pgelement{r_2}}) := \pglabellit{rider}
\ \ \ \
\pglambda({\pgelement{d_1}}) := \pglambda({\pgelement{d_2}}) := \pglabellit{driver} \\
& \pglambda({\pgelement{i_1}}) := \pglambda({\pgelement{i_2}}) := \pglabellit{pickup}
\ \ \ \
\pglambda({\pgelement{o_1}}) := \pglabellit{dropoff} \\
& \pglambda({\pgelement{s_1}}) := \pglambda({\pgelement{s_3}}) := \pglabellit{pickupTime}
\ \ \ \
\pglambda({\pgelement{s_2}}) := \pglabellit{dropoffTime} \\
& \pgupsilon(\pgelement{u_1}) := \pgupsilon(\pgelement{u_2}) := \pgupsilon(\pgelement{u_3}) := \pgupsilon(\pgelement{t_1}) := \pgupsilon(\pgelement{t_2}) := \pgupsilon(\pgelement{p_1}) := \pgupsilon(\pgelement{p_2}) := \pgupsilon(\pgelement{p_3}) := \ \pgunit \\
& \pgupsilon(\pgelement{r_1}) := (\pgelement{t_1}, \pgelement{u_2})
\ \ \ \
\pgupsilon(\pgelement{r_2}) := (\pgelement{t_2}, \pgelement{u_3})
\ \ \ \
\pgupsilon(\pgelement{d_1}) := (\pgelement{t_1}, \pgelement{u_1})
\ \ \ \
\pgupsilon(\pgelement{d_2}) := (\pgelement{t_2}, \pgelement{u_1}) \\
& \pgupsilon(\pgelement{i_1}) := (\pgelement{t_1}, \pgelement{p_1})
\ \ \ \
\pgupsilon(\pgelement{i_2}) := (\pgelement{t_2}, \pgelement{p_2})
\ \ \ \
\pgupsilon(\pgelement{o_1}) := (\pgelement{t_1}, \pgelement{p_2}) \\
& \pgupsilon(\pgelement{s_1}) := (\pgelement{i_1}, \pgvaluelit{1602203601})
\ \ \ \
\pgupsilon(\pgelement{s_2}) := (\pgelement{o_1}, \pgvaluelit{1602203948})
\ \ \ \
\pgupsilon(\pgelement{s_3}) := (\pgelement{i_2}, \pgvaluelit{1602204122})
\end{align*}
\end{footnotesize}

In general, we can easily classify property graph elements in terms of schema structure:
\begin{itemize}
    \item {\bf Vertex}: an element $\pgelement{e}$ for which $(\pgsigma \circ \pglambda)(\pgelement{e}) = \pgtypelit{1}$. The label $\pglambda(\pgelement{e})$ is said to be a {\it vertex label}.
    
    \item {\bf Edge}: an element $\pgelement{e}$ for which $(\pgsigma \circ \pglambda)(\pgelement{e}) = \pglabel{v_1} \times \pglabel{v_2}$, with $\pglabel{v_1}$ and $\pglabel{v_2}$ both vertex labels
    
    \item {\bf Vertex property}: an element $\pgelement{e}$ for which $(\pgsigma \circ \pglambda)(\pgelement{e}) = \pglabel{v} \times \pgtype{t}$, where $\pglabel{v}$ is a vertex label and $\pgtype{t}$ is any type whose definition does not contain a label; a property connects an element to a literal value, never to another element. Similarly:
    
    \item {\bf Edge property}: analogous to a vertex property; replace $\pglabel{v}$ with $\pglabel{e}$, where $\pglabel{e}$ is an edge label rather than a vertex label.
    
    \item {\bf Vertex meta-property}: analogous to a vertex or edge property, but the element label is a vertex property label rather than a vertex or edge label. Meta-properties are ``properties on properties'' and are supported by some but not all property graph implementations.
\end{itemize}

Other, less common classes of elements, such as edge meta-properties, and meta-edges (edges between edges or properties) can be defined similarly, but are very rarely seen in practice.
One can also relax the above constraints in various ways, giving rise to edge labels which admit a union of vertex labels on the head or tail of the edge, for example.

\subsection{Streams}
\label{section.streams}

Finally, let us consider an example of streams as APGs.
The following is a schema for streams:
\begin{footnotesize}
\begin{align*}
& \pgL := \{ \pglabellit{A} \} \\
& \pgsigma(\pglabellit{A}) := \pglabellit{A}+\pgone 
\end{align*}
\end{footnotesize}

Now a finite stream:
\begin{footnotesize}
\begin{align*}
& \pgelement{E}(\pglabellit{A}) := \{ \pgelement{e_1},\pgelement{e_2},\pgelement{e_3},\ldots,\pgelement{e_{10}}\} \\
& \pgupsilon(\pgelement{e_n}) := {\sf inl}(\pgelement{e_{n+1}})\textrm{ for $n=1,\ldots,9$} \\
& \pgupsilon(\pgelement{e_{10}}) := {\sf inr}(\pgunit)
\end{align*}
\end{footnotesize}

and an infinite stream:
\begin{footnotesize}
\begin{align*}
& \pgelement{E}(\pglabellit{A}) := \{ \pgelement{e_1},\pgelement{e_2},\pgelement{e_3},\ldots,\pgelement{e_{10}}\} \\
& \pgupsilon(\pgelement{e_n}) := {\sf inl}(\pgelement{e_{n+1}})\textrm{ for $n=1,\ldots,9$} \\
& \pgupsilon(\pgelement{e_{10}}) := {\sf inl}(\pgelement{e_4})
\end{align*}
\end{footnotesize}
\section{Equivalent Formulations}
\label{section.connections}

\subsection{APGs as coalgebras}\label{coalgebras}

It is useful to notice that APGs are coalgebras.  Indeed, given an APG schema $S=(\pgL,\pgsigma)$, we obtain a functor $F_S:{\sf Set}^{\pgL}\to{\sf Set}^{\pgL}$ mapping the family $(\pgelement{E}(\pglabel{l}))_{\pglabel{l}\in\pgL}$ to the family$(\pgvalue{V}(\pgsigma(\pglabel{l})))_{\pglabel{l}\in\pgL}$, where $\pgvalue{V}$ is defined in terms of $\pgvalue{E}$ as in (\ref{VintermsofE}).  Then, by definition, an APG is a coalgebra for $F_S$.  Through this characterization we find that APGs on the schema $S$ form the category $F_S\sf{-Coalg}$, which we write $S\sf{-APG}$, and we also gain access to the growing body of results and intuition for coalgebras.

\subsection{APGs as models}\label{models}

Another fruitful viewpoint is that APGs are models of ``theories''.  Given an APG schema $S=(\pgL,\pgsigma)$, we construct the free category $C_S$ with terminal and initial object, products, and coproducts, on the following generators:\footnote{See \cite{seely} for the details of this construction.  Note that the objects of $C_S$ are exactly $\pgtype{\textrm{Ty}}(\pgL)$.}
\begin{align*}
&\frac{\pglabel{l}\in\pgL}{({\sf Lbl} \ \pglabel{l})\in C_S} \ \ \ \ 
\frac{\pgtype{p} \in {\pgprimitives}}{({\sf Prim} \ \pgtype{p})\in C_S} \ \ \ \ 
\frac{\pglabel{l}\in\pgL}{\pgvalue{\delta}_{\pglabel{l}}:\pglabel{l}\to\pgsigma(\pglabel{l})}
\end{align*}

We call the category $C_S$ the {\it APG theory} corresponding to the APG schema $S$.  We then define a model of $C_S$ to be a functor $C_S\to{\sf Set}$ which preserves products, coproducts, terminal object, and initial object, and which sends ${\sf Prim} \ \pgtype{p} \mapsto {\pgprimvals}(\pgtype{p})$.
Now given an APG $(\pgelement{E},\pgupsilon)$ on $S$, we have a unique model $\pgvalue{M}:C_S\to {\sf Set}$ defined by
\begin{align*}
\pgvalue{M}(\pgtype{t})&:=\pgvalue{V}(\pgtype{t}) \\ \pgvalue{M}(\pgvalue{\delta}_{\pglabel{l}})&:=\pgupsilon_{\pglabel{l}}
\end{align*}
Conversely, a model $\pgvalue{M}$ of $C_S$ induces an APG on $S$ by defining
\begin{align*}
\pgelement{E}(\pglabel{l})&:=\pgvalue{M}({\sf Lbl} \ \pglabel{l})\\
\pgupsilon_{\pglabel{l}}&:=\pgvalue{M}(\pgvalue{\delta}_{\pglabel{l}})
\end{align*}
A morphism of models $\pgvalue{M}$ and $\pgvalue{M'}$ on $C_S$ is defined as a natural transformation $\eta:\pgvalue{M}\Rightarrow\pgvalue{M'}$ such that
\begin{align*}
&\eta_{{\sf Prim} \ \pgtype{p}} = {\sf id}_{{\sf Prim} \ \pgtype{p}} \ \ \ \ 
\eta_{\pgtype{t}+\pgtype{t'}} = \eta_{\pgtype{t}}+\eta_{\pgtype{t'}} \ \ \ \ 
\eta_{\pgtype{t}\times\pgtype{t'}} = \eta_{\pgtype{t}}\times \eta_{\pgtype{t'}}
\end{align*}
and the resulting category of models on $C_S$ is equivalent to $F_S\sf{-Coalg}$.
We also could choose to require that in $C_S$, the canonical maps $\pgtype{t}\times \pgtype{u} + \pgtype{t}\times \pgtype{v}\to \pgtype{t}\times (\pgtype{u}+\pgtype{v})$ and $\pgtype{t}\to \pgtype{t}+\pgzero$ are isomorphisms, making $C_S$ a distributive category.  This has no effect on the models of $C_S$, since ${\sf Set}$ is distributive, but it does allow for more schema morphisms (see \autoref{sec:morphisms}).
The downside is that it complicates term rewriting.

\section{Extensions}
\label{section.extensions}

\subsection{User-defined functions and constraints}

In many scenarios, we would like the ability to apply user-defined functions (UDFs) to individual data values, such as a function to convert strings to lowercase, or to compute the distance of latitude/longitude pairs from a reference point.  To obtain this ability, we add additional generators ${\sf f} : \pgtype{t} \to \pgtype{t'}$ to our APG theory $C_S$.
We also often would like to enforce constraints on APGs, for example we might want to constrain user-defined functions $\texttt{plus}$ and $\texttt{neg}$ so that $\texttt{plus}(\pgvaluelit{(1,2)}) = \pgvaluelit{3} = \texttt{neg}(\pgvaluelit{-3})$.  We can do this simply by adding the constraints as additional generating equalities to $C_S$.
In practice, we can implement these UDFs in languages like Java or Python instead of specifying an exhaustive set of axioms (which may be infinite), similarly to~\cite{wadt}, and we typically disallow labels in the types of UDFs, as such functions are not necessarily portable between APG schemas.

\subsection{Further type system extensions}
\label{section.moretypes}
Extending the APG type system to handle the more complex types encountered in practice is an important direction for future work.  To add function types, we can consider the free bi-cartesian \textit{closed} category on $B$, which adds exponential objects/types and $\lambda$-terms (or their combinator equivalents, ${\sf curry}$ and ${\sf apply}$) to $\mathcal{A}_B$~\cite{gabriel}.
To add inductive data types such as finite lists or finite trees, we can instead consider bi-cartesian categories that admit initial algebras for polynomial endofunctors, which contain a ``{\sf fold}'' operation that expresses structural recursion~\cite{grust}.  Dually, to add co-inductive data types such as infinite streams, we can consider bi-cartesian categories that admit final co-algebras for polynomial endo-functors, which contain an ``{\sf unfold}'' operation that expresses structural co-recursion. 
To add (not necessarily inductive) collection types such as sets and bags, {\it collection monads} can be added to our type theory, along with associated comprehension syntax (or monad combinators, {\sf return, map, bind}) for forming nested collections~\cite{grust}.
Such complex objects are common in schemas at Uber (see \autoref{section.implementations}) though they are usually avoided in typical property graph applications by reifying collections of elements using edges.

For example, a linked list of $\pglabellit{User}$ vertices can be realized in property graphs using something like a $\pglabellit{UserList}$ vertex label together with two edge labels: $\pglabellit{first} : \pglabellit{UserList} \rightarrow \pglabellit{User}$ and $\pglabellit{rest} : \pglabellit{UserList} \rightarrow \pglabellit{UserList}$.
Recursive types are not even possible in APG unless meta-edges
(see \autoref{section.examples})
are allowed; however, when they are, it is easy to see that APG can encode all recursive algebraic datatypes (lists, trees, natural numbers, etc.)~\cite{mitchell}.

\section{Morphisms of APG schemas}
\label{sec:morphisms}
\label{sec:ops}
We conclude our formalization of algebraic property graphs by briefly studying morphisms of APG schemas and how they can be applied to an APG.  The algorithms in this section are implemented, along with all of the examples in this paper, in the CQL tool, and Coq~\cite{Bertot:2010:ITP:1965123} proofs of all the theorems in this paper are also available.\footnote{\url{http://categoricaldata.net/APG.v}} We begin with preliminaries: structural recursion on APG types and terms, and then show how APG schemas can be read as E/R diagrams, then we describe morphisms of APGs on a given schema and morphisms of schemas, and conclude by showing how to transform APGs from one schema to another.

\subsection{Definition via APG theories}\label{theorymorphisms}

The clearest way to define morphisms of APG schemas is through their corresponding APG theories (see \autoref{models}).
Indeed, given APG schemas $S$ and $S'$, we form the APG theories $C_S$ and $C_{S'}$ (see \autoref{models}).
Then we define a morphism from $S$ to $S'$ simply as a functor $\Phi:C_S\to C_{S'}$ preserving products, coproducts, terminal object, initial object, and primitive types.

We can apply this morphism to an APG on $S'$ to obtain an APG on $S$.  To do this, we simply compose $\Phi$ with the corresponding model $\pgvalue{M'}:S'\to {\sf Set}$ to obtain a model $\Delta_{\Phi}(\pgvalue{M'}):=(S\xrightarrow{\Phi}S'\xrightarrow{\pgvalue{M'}}{\sf Set})$.
In this way, we have a category ${\sf Schema}$ of APG schemas and a contravariant functor ${\sf Schema}^{\textrm{op}}\to{\sf Cat}$ sending $S$ to the category $S{\sf -APG}$ of APGs on $S$ and sending $\Phi:S\to S'$ to the functor $\Delta_{\Phi}:S'{\sf -APG}\to S{\sf -APG}$.
The data migrations expressible as $\Delta_{\Phi}$ for some morphism $\Phi$ of APG schemas include dropping labels, duplicating labels, permuting the fields of product types, and, when we assume equality functions, joining of labels.  As a simple example, consider the schema $S=(\pgL,\pgsigma)$ with one label $\pglabel{l}$ with $\pgsigma(\pglabel{l}):=\pgtypelit{String} \times \pgtypelit{Nat} \times \pgtypelit{Integer}$ and the schema $S'=(\pgLprime,\pgsigma')$ with one label $\pglabel{l'}$ with $\pgsigma'(\pglabel{l'}):=\pgtypelit{Nat} \times \pgtypelit{String}$.

Then an example schema mapping $\Phi : C_S \to C_{S'}$ sends $\pglabel{l}$ to $\pglabel{l'}$ and $\pgvalue{\delta}_{\pglabel{l}}$ to the composite morphism $\pglabel{l'}\xrightarrow{\pgvalue{\delta}_{\pglabel{l'}}} \pgtypelit{Nat} \times \pgtypelit{String} \xrightarrow{\langle{\sf snd}, {\sf fst}, \pgvalue{42}\circ ! \rangle} \pgtypelit{String} \times \pgtypelit{Nat} \times \pgtypelit{Integer}$ (here we assume a user-defined function $\pgvalue{42}:\pgone\to\pgtypelit{String}$ sending $\pgunit\mapsto\pgvalue{42}$).  The functor $\Delta_{\Phi}$ converts APGs on $S'$ to schema $S$ by permuting projections and adding $\pgvalue{42}$.  

As a more sophisticated example of a morphism of APG schemas, we describe a morphism from the schema in Example~\ref{section.streams} to itself.  In this example we assume that $C_S$ is constructed to be distributive (see the note at the end of Subsection~\ref{models}).
\begin{align*}
\Phi(\pglabel{A}) &:= \pglabel{A}\times\pglabel{A} \\
\Phi(\pgvalue{\delta}_{\pglabel{A}}) &: \pglabel{A}\times\pglabel{A} \to \pglabel{A}\times\pglabel{A} + \pgone \\ 
\Phi(\pgvalue{\delta}_{\pglabel{A}}) &:= \Big{(} \pglabel{A}\times\pglabel{A} \xrightarrow{\pgvalue{\delta}_{\pglabel{A}} \times \pgvalue{\delta}_{\pglabel{A}}} (\pglabel{A}+\pgone)\times(\pglabel{A}+\pgone) \xrightarrow{(\pgvalue{\delta}_{\pglabel{A}}+\pgone) \times {\sf id}}
((\pglabel{A}+\pgone)+\pgone)\times(\pglabel{A}+\pgone) \\ &\xrightarrow{\textrm{dist},\textrm{assoc}} \pglabel{A}\times\pglabel{A} + (\pgone+\pgone)\times\pglabel{A} + \pglabel{A}\times \pgone + (\pgone+\pgone)\times \pgone \xrightarrow{[{\sf id},!,!,!]} \pglabel{A}\times\pglabel{A} + \pgone \Big{)}
\end{align*}

We apply this morphism to the finite stream $(\pgelement{E},\pgupsilon)$ in Example~\ref{section.streams} to obtain the following result:
\begin{footnotesize}
\begin{align*}
\Delta_{\Phi}(\pgelement{E})(\pglabel{A}) &:= \{ (\pgelement{e_i},\pgelement{e_j})\ |\ i,j=1,\ldots,10 \}\\
\Delta_{\Phi}(\pgelement{E})(\pgupsilon)((\pgelement{e_i},\pgelement{e_j})) &:= {\sf inl}((\pgelement{e_{i+2}},\pgelement{e_{j+1}}))\textrm{ for $i=1,\ldots,8$ and $j=1,\ldots,9$} \\
\Delta_{\Phi}(\pgelement{E})(\pgupsilon)((\pgelement{e_i},\pgelement{e_{10}})) &:= {\sf inr}(\pgunit)  \textrm{ for $i=1,\ldots,10$} \\
\Delta_{\Phi}(\pgelement{E})(\pgupsilon)((\pgelement{e_i},\pgelement{e_j})) &:= {\sf inr}(\pgunit)  \textrm{ for $i=9,10$ and $j=1,\ldots,10$}
\end{align*}
\end{footnotesize}
We can say that $\Delta_{\Phi}$ converts a stream into a new stream whose elements are ordered pairs of elements of the old stream, and where the first element advances twice as fast as in the old stream, and the second element advances just as in the old stream.


\subsection{Definition using coalgebra interpretation}

Using the coalgebra interpretation (see Subsection~\ref{coalgebras}), we are tempted to define a morphism between APG schemas $S=(\pgL,\pgsigma)$ and $S'=(\pgLprime,\pgsigmaprime)$ as follows.  We first consider the corresponding functors $F_S:{\sf Set}^{\pgL}\to {\sf Set}^{\pgL}$ and $F_{S'}:{\sf Set}^{\pgLprime}\to {\sf Set}^{\pgLprime}$. Then we might think to define a morphism as a pair $(\Phi:{\sf Set}^{\pgLprime}\to {\sf Set}^{\pgL},\ \phi:\Phi F_{S'}\Rightarrow F_S\Phi)$, as shown:
\[
\begin{tikzcd}
{\sf Set}^{\pgLprime} \ar[r, "F_{S'}"] \ar[d, "\Phi"] & {\sf Set}^{\pgLprime} \ar[d, "\Phi"]  \ar[ld, Rightarrow, shorten <>=4pt, "\phi"] \\
{\sf Set}^{\pgL} \ar[r, "F_S"] & {\sf Set}^{\pgL}
\end{tikzcd}
\]
We can apply this pair to an APG $(\pgelement{E'}:\pgLprime\to {\sf Set}, \pgupsilonprime:\pgelement{E'}\Rightarrow F_{S'}(\pgelement{E'}))$ to obtain an APG
\begin{align*}
    \pgelement{E}&:=\Phi(\pgelement{E'})\\
    \pgupsilon&:=\Big{(}\Phi(\pgelement{E'})\xRightarrow{\Phi(\pgupsilonprime)}\Phi F_{S'}(\pgelement{E'})\xRightarrow{\phi_{\pgelement{E'}}}F_S\Phi(\pgelement{E'})\Big{)}
\end{align*}
For the purpose of implementation, we must limit the possibilities for $\Phi$ somehow, or we will include non-computable transformations.  A reasonable condition would be that it is of the form $(\pgelement{E'}(\pglabel{l'}))_{\pglabel{l'}\in\pgLprime}\mapsto (\pgvalue{V'}(f(\pglabel{l})))_{\pglabel{l}\in\pgL}$ (where $\pgvalue{V'}$ is defined in terms of $\pgelement{E'}$ as in Section~\ref{section.data-model}) for some $f:\pgL\to \pgtype{\textrm{Ty}}(\pgLprime)$, just as $F_S$.  With this restriction, we obtain a strictly smaller class of morphisms than described in Subsection~\ref{theorymorphisms}.  In fact, the example above with streams can be verified not to be a morphism of this type.
However, we think that it is possible to modify this definition so that it coincides with the definition above.  We do this in a sequence of conjectures which we think are true --- proofs are forthcoming.\footnote{These conjectures, as well as the next section, were conceived partly in conversation with David Spivak.}

First consider the category ${\sf Comonad}$ whose objects are comonads $(A,D:A\to A, \varepsilon:D\Rightarrow {\sf id}_A, \delta:D\Rightarrow D\circ D)$ and whose morphisms $(A',D',\varepsilon',\delta')\to (A,D,\varepsilon,\delta)$ are pairs $(F:A'\to A,f:F\circ D'\Rightarrow D\circ F)$ respecting the comonad structure as in \cite{street_1972}.

We also consider the category ${\sf LaxEnd}$ whose objects are endofunctors $(A,D:A\to A)$ and whose morphisms are again pairs $(F:A'\to A,f:F\circ D'\Rightarrow D\circ F)$.  Clearly there is a forgetful functor $U:{\sf Comonad}\to {\sf LaxEnd}$.

\begin{conjecture}
The functor $U$ has a partial right adjoint ${\sf Cof}:{\sf LaxEnd}\to {\sf Comonad}$ defined on the endofunctors we are considering, $({\sf Set}^{\pgL}, F_S)$.  Explicitly, $(F_S)^{\ast}:={\sf Cof}(F_S)$ sends $(\pgelement{E})$ to the terminal coalgebra of the endofunctor $\pgelement{E}\times F_S$.  The category of comonad coalgebras of ${\sf Cof}(F_S)$ is equivalent to the category of endofunctor coalgebras of $F_S$.
\end{conjecture}

Now we define a morphism as a pair $(\Phi:{\sf Set}^{\pgLprime}\to {\sf Set}^{\pgL},\ \phi:\Phi (F_{S'})^\ast\Rightarrow F_S\Phi)$:
\[
\begin{tikzcd}
{\sf Set}^{\pgLprime} \ar[r, "(F_{S'})^\ast"] \ar[d, "\Phi"] & {\sf Set}^{\pgLprime} \ar[d, "\Phi"]  \ar[ld, Rightarrow, shorten <>=4pt, "\phi"] \\
{\sf Set}^{\pgL} \ar[r, "F_S"] & {\sf Set}^{\pgL}
\end{tikzcd}
\]
In other words, we use co-Kleisli morphisms of the comonad $U\circ {\sf Cof}$.
To apply such a morphism to the APG $(\pgelement{E'}:\pgLprime\to {\sf Set}, \pgupsilonprime:\pgelement{E'}\Rightarrow F_{S'}(\pgelement{E'}))$, we first promote $\pgupsilonprime$ to a coalgebra $\overline{\pgupsilonprime}:\pgelement{E'}\Rightarrow (F_{S'})^{\ast}(\pgelement{E'}))$ of the comonad $(F_{S'})^{\ast}$.  Thence we obtain an APG on $S$:
\begin{align*}
    E&:=\Phi(\pgelement{E'})\\
    \pgupsilon&:=\Big{(}\Phi(\pgelement{E'})\xRightarrow{\Phi\left(\overline{\pgupsilonprime}\right)}\Phi (F_{S'})^{\ast}(\pgelement{E'})\xRightarrow{\phi_{\pgelement{E'}}}F_S\Phi(\pgelement{E'})\Big{)}
\end{align*}

\begin{conjecture}
These morphisms are equivalent to the morphisms in \autoref{theorymorphisms}, and have identical action on APGs.
\end{conjecture}

\section{Connection to Poly, and generalized APGs}
\label{section.poly}

As we have seen in Subsection~\ref{coalgebras}, an APG schema $S=(\pgL,\pgsigma)$ is a polynomial functor from ${\sf Set}^{\pgL}$ to itself.  But not all such polynomial functors are APG schemas.  For example, the polynomial functor $P:{\sf Set}^{\{\pglabel{l}\}}\to {\sf Set}^{\{\pglabel{l}\}}$ defined by $\pgelement{E}_{\pglabel{l}}\mapsto 1+\pgelement{E}_{\pglabel{l}}+\pgelement{E}_{\pglabel{l}}^2+\pgelement{E}_{\pglabel{l}}^3+\cdots$ is not an APG schema since it involves infinitary type constructors.

In the language of David Spivak~\cite{spivak2022polynomial}, these polynomial functors can alternatively be described as bicomodules in $\mathsf{Poly}$ from a discrete category (aka comonoid) to itself: $\pglabel{\mathcal{L}}y\bimodfrom[\pgtype{\sigma}]\pglabel{\mathcal{L}}y$.  We immediately generalize to the case of an arbitrary category $\pglabel{\mathcal{L}}$ of labels, to obtain a \textit{generalized APG (GAPG) schema}, defined as a category $\pglabel{\mathcal{L}}$ and a bicomodule (aka prafunctor) $\pglabel{\mathcal{L}}\bimodfrom[\pgtype{\sigma}]\pglabel{\mathcal{L}}$.
A \textit{GAPG} on the schema $(\pglabel{\mathcal{L}},\pgtype{\sigma})$ is then an ordered pair $(\pgelement{E},\pgupsilon)$ where $\pgelement{E}$ is a copresheaf on $\pglabel{\mathcal{L}}$, aka a bicomodule $\pglabel{\mathcal{L}}\bimodfrom[\pgelement{E}] 0$, and $\pgupsilon$ is a bicomodule homomorphism:
\[
\begin{tikzcd}
	\pglabel{\mathcal{L}}\ar[swap, rd, bimr-biml, "\pgtype{\sigma}"] \ar[rr,bimr-biml,"\pgelement{E}"] & \phantom{I} \ar[d,shorten <>=4pt,Rightarrow,"\pgupsilon"] & 0 \\
	& \pglabel{\mathcal{L}} \ar[ru,swap, bimr-biml, "\pgelement{E}"]
\end{tikzcd}
\]

We can then define a schema morphism $(\pglabel{\mathcal{L}},\pgtype{\sigma})\to (\pglabel{\mathcal{L}'},\pgtype{\sigma'})$ as an ordered pair $(\Psi,\psi)$:

\[
\begin{tikzcd}[sep=40pt]
	\pglabel{\mathcal{L}}\ar[r, bimr-biml, "\pgtype{\sigma}"] \ar[d, bimr-biml, "\Psi"] & \pglabel{\mathcal{L}} \ar[d, bimr-biml, "\Psi"] \\
	\pglabel{\mathcal{L}'}\ar[r, bimr-biml, "(\pgtype{\sigma'})^\ast"] \ar[ru,shorten <>=4pt,Rightarrow,"\psi"] & \pglabel{\mathcal{L'}}
\end{tikzcd}
\]

where $(\pgtype{\sigma'})^{\ast}$ is the cofree comonad on $\pgtype{\sigma'}$.
Assuming that the above conjectures continue to hold in this general setting, we can uniquely lift a GAPG $(\pgelement{E},\pgupsilon)$ to a right comodule $(\pgelement{E},\overline{\pgupsilon})$ of $(\pgtype{\sigma})^{\ast}$:

\[
\begin{tikzcd}
	\pglabel{\mathcal{L}}\ar[swap, rd, bimr-biml, "(\pgtype{\sigma})^{\ast}"] \ar[rr,bimr-biml,"\pgelement{E}"] & \phantom{I} \ar[d,shorten <>=4pt,Rightarrow,"\overline{\pgupsilon}"] & 0 \\
	& \pglabel{\mathcal{L}} \ar[ru,swap, bimr-biml, "\pgelement{E}"]
\end{tikzcd}
\]

Then the schema morphism $(\Psi,\psi)$ can be applied to a GAPG $(\pgelement{E'},\pgupsilonprime)$ on $(\pgLprime, \pgtype{\sigma'})$ in a straightforward way:

\[
\begin{tikzcd}[sep=60pt]
	\pglabel{\mathcal{L}}\ar[r, bimr-biml, "\pgtype{\sigma}^\ast"] \ar[d, bimr-biml, "\Psi"] & \pglabel{\mathcal{L}} \ar[d, bimr-biml, "\Psi"] \\
	\pglabel{\mathcal{L}'}\ar[r, bimr-biml, "(\pgtype{\sigma'})^\ast"{name=monad}] \ar[ru,shorten <>=4pt,Rightarrow,"\psi"] \ar[dr, bimr-biml, swap, "\pgelement{E'}"{name=X}] & \pglabel{\mathcal{L'}} \ar[d, bimr-biml, "\pgelement{E'}"] \ar[from=X, Rightarrow, shorten <>=4pt, "\overline{\pgupsilonprime}"] \\
	& 0
\end{tikzcd}
\]

Use cases and implementation possibilities of GAPGs have yet to be explored.

\section{Implementations}
\label{section.implementations}


\begin{wrapfigure}{r}{0.15\textwidth}
    \centering
    \includegraphics[width=0.15\textwidth]{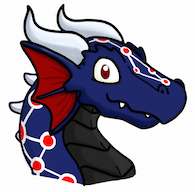}
\end{wrapfigure}
Dragon\footnote{\url{https://eng.uber.com/dragon-schema-integration-at-uber-scale}} is a logical data model and a proprietary framework for data and schema transformations at Uber, with overlapping implementations in Haskell and Java.
It is most often used for carrying standardized data type definitions into each schema language used at Uber, for validating schemas in a number of languages, for migrating schemas between languages, and even for migrating data and schemas in parallel.
For example, daily snapshots of Uber's metadata catalog\footnote{\url{https://eng.uber.com/metadata-insights-databook}} may be converted from JSON to RDF at the same time as its schemas are converted from YAML to SHACL.
Protocol Buffers messages are converted to Avro records in a streaming fashion while maintaining consistency with a one-time mapping of the Protobuf schemas to Avro schemas.
Provided that data ``on the left'' conforms to a schema ``on the left'', the transformed data ``on the right'' is guaranteed to conform to the transformed schema ``on the right''.
The framework is also used for statistical and topological analysis of Uber's many thousands of schemas, and for applying common ``linter'' rules and best practices to these schemas in parallel, in a language-neutral way.
Dragon takes its name from a discussion at Dagstuhl Seminar 19419~\cite{sakr2021future} about the best way to build composable mappings among multiple data models: in a pairwise fashion, or in a star configuration with a potentially complex ``dragon'' data model in the center.

A complete implementation of APG, albeit without the mappings to external data models and additional features of Dragon, is also provided with CQL.

Finally, Hydra\footnote{\url{https://github.com/CategoricalData/hydra}} is a new open source framework currently under development which is similar to, but more advanced than Dragon.
Hydra extends the type grammar provided in \autoref{section.data-model} to include all of System F, and supports Hindley-Milner type inference.
The new framework is designed to carry not only schemas and algebraic data, but also purely functional program code between languages.

\section{Conclusion}
\label{section.summary}

In this paper, we have provided a sound mathematical basis for a family of data models we call algebraic property graphs, representing a bridge between heavily used graph and non-graph data models, helping to broaden the scope of graph computing and lower the barrier to building enterprise knowledge graphs at scale.
The paper has presented multiple points of view on APGs: they can be considered as coalgebras of an endofunctor, models of a theory, or generalized to GAPGs in the framework of ${\sf Poly}$.
Each of these points of view brings different aspects of APGs into focus, as well as connecting them to the large bodies of literature on coalgebra, categorical logic, and ${\sf Poly}$ respectively.
The conjectures in the paper attempt to relate these points of view in the context of data migration, but more work is needed to adequately liaise between this viewpoints.

Among many possible ways of standardizing the popular notion of a property graph, we believe the use of algebraic data types is especially promising due to their ubiquity and conceptual simplicity.
In addition, the details of the relationship between algebraic property graphs and algebraic databases~\cite{wadt} is a promising line of future work with immediate applications in industry.


\bibliography{main}






\end{document}